\documentclass[conference]{IEEEtran}
\IEEEoverridecommandlockouts

\IEEEpubid{
\begin{minipage}{\textwidth}\ \\[8pt]
\centering
  \copyright~2026 IEEE.  Personal use of this material is permitted.  Permission from IEEE must be obtained for all other uses, in any current or future media, including reprinting/republishing this material for advertising or promotional purposes, creating new collective works, for resale or redistribution to servers or lists, or reuse of any copyrighted component of this work in other works. 
\end{minipage}
}

\usepackage{cite}
\usepackage{amsmath,amssymb,amsfonts}
\usepackage{algorithmic}
\usepackage{graphicx}
\usepackage{textcomp}
\usepackage{xcolor}
\usepackage{placeins}
\usepackage{array}
\usepackage{booktabs}   
\usepackage{multirow}   
\usepackage{makecell}   
\usepackage{graphicx}
\usepackage{adjustbox}
\usepackage{subcaption}               
\usepackage{lipsum}

\def\BibTeX{{\rm B\kern-.05em{\sc i\kern-.025em b}\kern-.08em
    T\kern-.1667em\lower.7ex\hbox{E}\kern-.125emX}}
\begin{document}

\title{Task-Specific Sparse Feature Masks for Molecular Toxicity Prediction with Chemical Language Models
\thanks{This work is supported by Hong Kong Metropolitan University (Project RD/2025/1.15) and Hong Kong Research Grants Council (Project UGC/FDS16/E16/23).}
}

\author{
\IEEEauthorblockN{Kwun Sy Lee}
\IEEEauthorblockA{\textit{School of Science and Technology} \\
\textit{Hong Kong Metropolitan University}\\
Ho Man Tin, Hong Kong\\
kwslee@hkmu.edu.hk}
\and

\IEEEauthorblockN{Jiawei Chen}
\IEEEauthorblockA{\textit{School of Science and Technology} \\
\textit{Hong Kong Metropolitan University}\\
Ho Man Tin, Hong Kong\\
jwchen@hkmu.edu.hk}
\and

\IEEEauthorblockN{Fuk Sheng Ford Chung}
\IEEEauthorblockA{\textit{School of Science and Technology} \\
\textit{Hong Kong Metropolitan University}\\
Ho Man Tin, Hong Kong\\
ffschung@hkmu.edu.hk}
\and

\IEEEauthorblockN{Tianyu Zhao}
\IEEEauthorblockA{\textit{Faculty of Engineering} \\
\textit{Hong Kong Polytechnic University}\\
Hung Hom, Hong Kong \\
ztyshawnnn@gmail.com}
\and

\IEEEauthorblockN{Zhenyuan Chen}
\IEEEauthorblockA{\textit{School of Science and Technology} \\
\textit{Hong Kong Metropolitan University}\\
Ho Man Tin, Hong Kong\\
chenzh@hkmu.edu.hk}
\and

\IEEEauthorblockN{Debby D. Wang\IEEEauthorrefmark{1}}
\IEEEauthorblockA{\textit{School of Science and Technology} \\
\textit{Hong Kong Metropolitan University}\\
Ho Man Tin, Hong Kong\\
dwang@hkmu.edu.hk \\
Corresponding author\IEEEauthorrefmark{1}}
}

\maketitle

\begin{abstract}

Reliable \textit{in silico} molecular toxicity prediction is a cornerstone of modern drug discovery, offering a scalable alternative to experimental screening. However, the black-box nature of state-of-the-art models remains a significant barrier to adoption, as high-stakes safety decisions demand verifiable structural insights alongside predictive performance. To address this, we propose a novel multi-task learning (MTL) framework designed to jointly enhance accuracy and interpretability. Our architecture integrates a shared chemical language model with task-specific attention modules. By imposing an L1 sparsity penalty on these modules, the framework is constrained to focus on a minimal set of salient molecular fragments for each distinct toxicity endpoint. The resulting framework is trained end-to-end and is readily adaptable to various transformer-based backbones. Evaluated on the ClinTox, SIDER, and Tox21 benchmark datasets, our approach consistently outperforms both single-task and standard MTL baselines. Crucially, the sparse attention weights provide chemically intuitive visualizations that reveal the specific fragments influencing predictions, thereby enhancing insight into the model's decision-making process. 

\end{abstract}


\section{Introduction}

The assessment of molecular toxicity is a crucial yet challenging step in early-stage drug discovery and chemical safety screening. Traditional experimental assays are often resource-intensive and time-consuming, creating a significant bottleneck. Consequently, developing accurate and efficient \textit{in silico} methods has become a key priority in computational toxicology to accelerate the identification of potentially hazardous compounds.

\IEEEpubidadjcol

In recent years, the field has increasingly adopted sequence-based deep learning approaches operating on SMILES representations of molecules. Chemical language models, such as ChemBERTa~\cite{chithrananda2020chemberta}, MoLFormer~\cite{ross2022large}, and SMILES-BERT~\cite{wang2019smiles}, pre-trained on vast unlabeled chemical corpora, have shown remarkable success in learning rich, contextual representations for downstream prediction tasks. However, their performance is often constrained by the scarcity of labeled toxicological data. To address this, MTL~\cite{caruana1997multitask} has become a widely adopted paradigm~\cite{ramsundar2015massively, allenspach2024neural}, enabling models to generalize better by jointly learning from several related endpoints. 

Despite their advantages, prevailing transformer-based MTL architectures~\cite{peng2020empirical, zhang2022pushing}, which typically use a hard parameter sharing (HPS)~\cite{caruna1993multitask} approach, face two critical limitations. First, HPS lacks an explicit mechanism for task-specific feature selection; the final latent representation from the shared backbone is passed wholesale to each prediction head, assuming all features are equally relevant to every downstream task. This is often suboptimal, as distinct toxicity mechanisms depend on different molecular properties~\cite{zhu2021genotype}, potentially leading to negative transfer where conflicting gradients destabilize the shared representation~\cite{ruder2017overview}. Second, the opacity of these models restricts their utility in safety-critical decision making. While standard architectures may achieve high predictive accuracy, they obscure the decision-making process, failing to provide transparent feature attribution relative to the underlying chemical structure. In toxicology, where adverse effects are frequently driven by precise molecular substructures (toxicophores), this interpretability gap remains a significant barrier to integration into discovery pipelines. 

To bridge this gap, we introduce a model-agnostic MTL framework that simultaneously enhances performance and interpretability via task-specific feature gating. Our contributions are three-fold: 
\begin{enumerate}
    \item We propose an architecture that adapts task-specific attention mechanisms (inspired by MTAN~\cite{liu2019end}) into an HPS framework. This design enables each toxicity endpoint to learn a tailored representation from a shared chemical embedding generated by any underlying transformer-based backbone. 
    \item We introduce an L1 sparsity penalty on the attention masks to constrain the model's focus to a minimal subset of salient molecular fragments for each task. This regularization aligns the model's feature selection process with the toxicological intuition that specific substructures often drive toxic responses. 
    \item We validate our model on benchmark toxicity datasets, demonstrating competitive predictive performance while generating chemically intuitive visualizations. These visualizations highlight molecular features that facilitate expert verification, thus enhancing the transparency of the model's decision-making. 
\end{enumerate}

\section{Methods}

We now introduce our MTL framework, dissect the task-specific attention modules, and describe the formulation of the objective function with sparsity regularization on the attention scores.

\subsection{Overall Architecture}

\begin{figure*}
    \centering
    \includegraphics[width=0.68\linewidth]{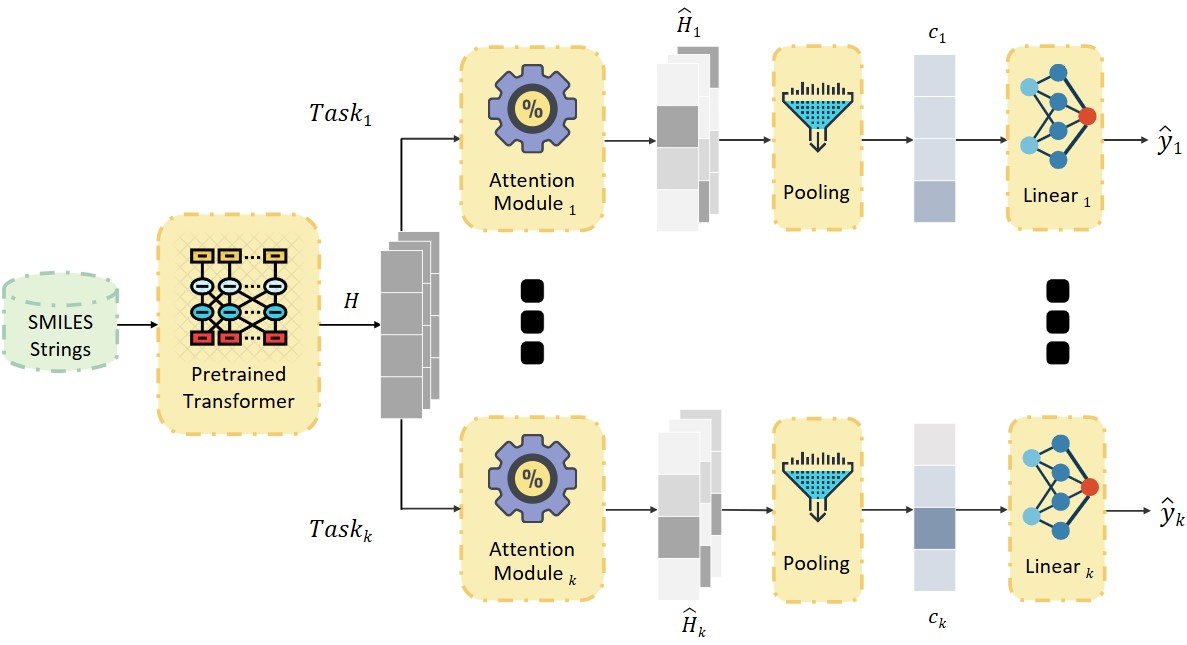}
    \caption{
    \textbf{Overall Multi-Task Learning Architecture.}
    A shared transformer backbone processes SMILES strings into a contextual chemical representation ($\mathbf{H}$). This representation is distributed to $K$ task-specific heads. In each head, an attention module (Fig.~\ref{fig:task_attn_module}) generates a task-specific feature set ($\hat{\mathbf{H}}_k$), which is then aggregated via weighted pooling into a fixed-size context vector ($\mathbf{c}_k$) for final prediction ($\hat{y}_k$). 
    }
    \label{fig:arch_overview}
\end{figure*}

Fig.~\ref{fig:arch_overview} provides an overview of our MTL architecture. The model is built upon a HPS paradigm, where a single, shared backbone is utilized across all tasks. In our implementation, this backbone consists of a pre-trained transformer that processes tokenized SMILES strings to generate contextualized molecular representations. 

Branching from this shared backbone are $K$ task-specific heads, each incorporating our proposed attention module. This module learns a task-specific soft attention mask that re-weights the shared representations from the backbone. Thus, while the backbone is trained to generate broadly applicable chemical features, the attention modules provide a mechanism to independently re-weight these representations for each toxicological endpoint. This design facilitates a joint optimization of shared knowledge and task-specific specialization. 

To generate the final prediction, the re-weighted features are aggregated into a single, fixed-size context vector for each task. This is achieved via a weighted pooling operation: the masked features are summed and subsequently normalized by the accumulated attention scores to maintain representation stability. The resulting vector is then fed into a task-specific linear layer to obtain the final output logit. 

\subsection{Task-Specific Attention Module}

\begin{figure}
    \centering
    \includegraphics[width=0.58\linewidth]{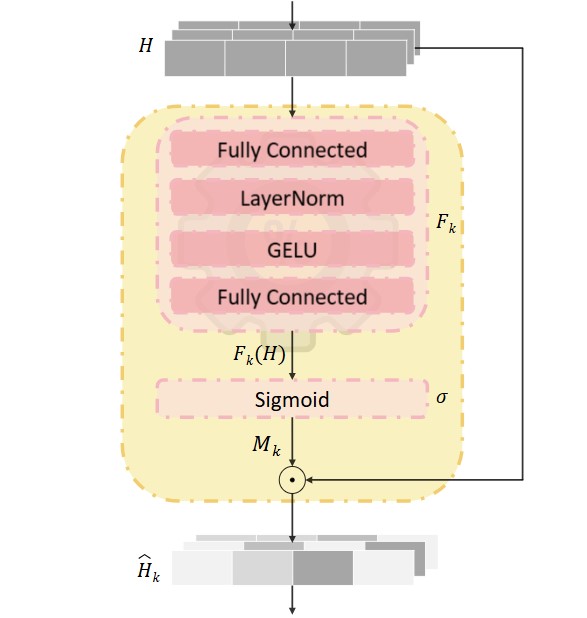}
    \caption{
    \textbf{Task-Specific Attention Module.} 
    Shared features ($\mathbf{H}$) are processed by a feed-forward network ($\mathit{F}_k$) and a Sigmoid activation ($\sigma$) to generate a soft mask ($\mathbf{M}_k$). This mask is applied to the original input via element-wise multiplication ($\odot$), producing a re-weighted, task-specific feature representation ($\hat{\mathbf{H}}_k$) where darker regions indicate higher salience.
    }
    \label{fig:task_attn_module}
\end{figure}

The core of each prediction head is a task-specific attention module, as depicted in Fig.~\ref{fig:task_attn_module}. Its purpose is to dynamically modulate the output of the shared backbone, creating a tailored feature representation for each task by suppressing irrelevant chemical tokens. 

Let the final hidden layer of the transformer backbone produce a sequence of token representations $\mathbf{H} \in \mathbb{R}^{L \times D_h}$, where $L$ is the sequence length and $D_h$ is the hidden dimension. Structurally, the attention module utilizes a lightweight sub-network to generate the mask scores. This sub-network is implemented as a two-layer feed-forward neural network (FFNN) with a LayerNorm and a GELU activation. Its output is passed through a sigmoid activation to ensure the attention weights are bounded between 0 and 1. 

The generation of the mask for task $k$, $\mathbf{M}_k = \{m_{k,1},\ldots,m_{k,L}\} \in [0, 1]^L$, can be expressed as: 
\begin{equation}
\mathbf{M}_k = \sigma(\mathit{F}_k(\mathbf{H})),
\label{eq.mask}
\end{equation}
where $\mathit{F}_k$ represents the task-specific FFNN and $\sigma$ is the sigmoid function. In the final gating step, this mask is then applied element-wise to the shared feature representation $\mathbf{H}$ to produce a task-specific feature set $\hat{\mathbf{H}}_k$:
\begin{equation}
\hat{\mathbf{H}}_k = \mathbf{M}_k \odot \mathbf{H},
\end{equation}
where $\odot$ denotes element-wise multiplication that is broadcast along the hidden dimension $D_h$. 

\subsection{Objective Function}
Our MTL objective is a composite function designed to optimize predictive accuracy while promoting interpretability by enforcing sparsity in the attention masks. The overall objective is constructed by first defining a total loss for each individual task, which includes both a prediction loss component and a regularization penalty. These task-level losses are then averaged to form the final training objective.

For each task $k$, the total loss $\mathcal{L}_k$ is given by:
\begin{equation}
\mathcal{L}_{k} = \mathcal{L}_{\text{pred},k} + \lambda \mathcal{R}_{\text{a},k},
\end{equation}
where $\mathcal{L}_{\text{pred},k}$ is the prediction loss, $\mathcal{R}_{\text{a},k}$ is the attention regularization term for that task, and $\lambda$ is a hyperparameter controlling the regularization strength.

The prediction component, $\mathcal{L}_{\text{pred},k}$, is the Binary Cross-Entropy (BCE) loss. For a dataset of $N$ molecules, it is computed as:
\begin{equation}
\mathcal{L}_{\text{pred},k} = \frac{1}{\sum_{n=1}^N \delta_{n,k}} \sum_{n=1}^N \delta_{n,k} \cdot \mathcal{L}_{\text{BCE}}( \sigma(\hat{y}_{n,k}), y_{n,k}),
\label{eq.pred_loss}
\end{equation}
where $y_{n,k}$ and $\hat{y}_{n,k}$ are the ground truth label and predicted logit for the $n$-th sample and $k$-th task, respectively, and $\sigma$ denotes the sigmoid function. The indicator $\delta_{n,k}$ is 1 if a valid label exists for sample $n$ on task $k$ and 0 otherwise, ensuring that the loss is computed only over labeled samples.

The regularization component, $\mathcal{R}_{\text{a},k}$, is an L1 penalty applied to the attention masks to encourage sparsity. This forces the model to focus on a minimal set of highly salient tokens, hence isolating the specific molecular fragments that most significantly influence the prediction. This penalty is averaged across all valid samples for task $k$:
\begin{equation}
\mathcal{R}_{\text{a},k} = \frac{1}{\sum_{n=1}^N \delta_{n,k}} \sum_{n=1}^N \delta_{n,k} \|\mathbf{M}_{n,k}\|_1,
\end{equation}
where $\mathbf{M}_{n,k}$ is the attention mask~\eqref{eq.mask} generated for the $n$-th sample by the $k$-th task's attention module.

The final, overall objective function $\mathcal{L}$ is the mean of these task-level losses:
\begin{equation}
\mathcal{L} = \frac{1}{K} \sum_{k=1}^K \mathcal{L}_{k} = \frac{1}{K} \sum_{k=1}^K \left( \mathcal{L}_{\text{pred},k} + \lambda \mathcal{R}_{\text{a},k} \right).
\end{equation}

\section{Experiments}

We evaluate our framework on molecular toxicity benchmarks to validate both predictive performance and interpretability. Our experiments span datasets of varying scale and task complexity. Specifically, we conduct experiments on three MoleculeNet~\cite{wu2018moleculenet} benchmarks: ClinTox (1,484 compounds, 2 tasks)~\cite{gayvert2016data}, SIDER (1,427 drugs, 27 adverse reaction tasks)~\cite{kuhn2016sider}, and Tox21 (7,831 compounds, 12 receptor assays)~\cite{huang2016modelling}. 

We instantiated three pre-trained transformer backbones via HuggingFace~\cite{wolf2019huggingface}. To assess performance across both specialized and general architectures, we selected ChemBERTa-1~\cite{chithrananda2020chemberta}, and MoLFormer~\cite{ross2022large} as representative chemical language models, while including DistilBERT~\cite{sanh2019distilbert} as a general-purpose NLP baseline. Our proposed MTL framework with sparse attention (MTL-SA) is compared against single-task learning (STL) and standard multi-task learning with hard parameter sharing (MTL-HPS) following~\cite{peng2020empirical, zhang2022pushing}. All models are trained using identical preprocessing and hyperparameter optimization procedures to ensure fair comparison. We first describe our experimental setup before presenting benchmark results, ablation studies on sparsity regularization, and qualitative analysis of learned attention patterns. 

\subsection{Experimental Setup}
\label{subsec:experimental_setup}

Raw SMILES strings are validated and canonicalized using RDKit. Each dataset is partitioned into 80\% training and 20\% test sets via iterative stratified sampling~\cite{sechidis2011stratification}. Hyperparameter optimization employs 5-fold cross-validation with Bayesian search using Optuna~\cite{akiba2019optuna}. Final models are retrained on the full 80\% training set and evaluated once on the held-out 20\% test set. Performance is measured using macro-averaged ROC-AUC across all tasks. 

All models are fine-tuned with AdamW~\cite{loshchilov2017decoupled} on NVIDIA A800 GPUs using PyTorch 2.6.0. 

\subsection{Results} 

This section presents benchmark performance comparisons across learning paradigms, followed by ablation studies examining the effect of L1 sparsity regularization strength.

\subsubsection{Benchmark Performance}

Table~\ref{tab:performance_comparison} reports the ROC-AUC scores across all learning frameworks. Classical machine learning models establish strong initial baselines, with Random Forest notably achieving a competitive score of 0.6581 on SIDER. As expected, transformer-based models generally deliver superior performance, particularly on the ClinTox and Tox21 datasets. However, standard MTL-HPS exhibits mixed efficacy; while it improves over STL in some cases, it leads to performance degradation in others (e.g., ChemBERTa-1 on ClinTox and SIDER), a phenomenon likely attributable to negative transfer. 

In contrast, our proposed method consistently outperforms both STL and MTL-HPS baselines across nearly all backbone-dataset combinations. The advantage is most pronounced when using specialized chemical language models, particularly the MoLFormer backbone, where our approach achieves the highest overall scores on ClinTox (0.9482) and Tox21 (0.8528), and secures the best average performance (0.8149) across all tested models. This consistent stability indicates that our task-specific gating implicitly mitigates negative transfer. By enabling each head to filter out irrelevant or conflicting features from the shared backbone, the model reduces the interference that typically hampers standard HPS architectures. 

\begin{table}[!htbp]
\centering
\caption{
    \textbf{Performance Comparison (ROC-AUC) on Benchmark Datasets.} Our proposed MTL-SA is compared against classical machine learning baselines, STL, and MTL-HPS across three transformer backbones. The highest score within each backbone-dataset group is marked in bold. The highest overall score achieved for each dataset across all methods is underlined.}
\label{tab:performance_comparison}

\resizebox{\columnwidth}{!}{
\begin{tabular}{@{} l c c c c c c @{}}
\toprule
\multirow{8}{*}{\textbf{Machine Learning}}& & & \multicolumn{4}{c}{\textbf{Performance}} \\
\cmidrule(lr){4-7}
& \multicolumn{2}{c}{\textbf{Model}}& (2 Tasks)& (27 Tasks) & (12 Tasks) & \\
& & & \textbf{ClinTox} & \textbf{SIDER} & \textbf{Tox21} & \textbf{Average} \\
\cmidrule{2-7}
& \multicolumn{2}{c}{ECFP+RF}& 0.8629 & \underline{\textbf{0.6581}} & \textbf{0.8199}& \textbf{0.7803}\\
& \multicolumn{2}{c}{ECFP+SVM}& \textbf{0.8657}& 0.6462 & 0.8057 & 0.7725 \\
& \multicolumn{2}{c}{ECFP+KNN}& 0.7079 & 0.5992 & 0.7300 & 0.6790 \\
\midrule[1.5pt]
 & \multirow{2}{*}{\textbf{Backbone}} & \textbf{Learning} & \multicolumn{4}{c}{\textbf{Performance}}\\
\cmidrule(lr){4-7}
& & \textbf{Framework} & \textbf{ClinTox} & \textbf{SIDER} & \textbf{Tox21} & \textbf{Average} \\
\cmidrule{2-7}
\multirow{9}{*}[1.5em]{\textbf{Transformers}}
& \multirow{3}{*}{DistilBERT}
& STL           & 0.9329 & 0.6036 & 0.8157 & 0.7841 \\
& & MTL-HPS       & 0.9360 & 0.6076 & 0.8419 & 0.7952 \\
& & MTL-SA (Ours) & \textbf{0.9370} & \textbf{0.6231} & \textbf{0.8439} & \textbf{0.8013} \\
\cmidrule{2-7}
& \multirow{3}{*}{ChemBERTa-1}
& STL           & 0.9202 & 0.6122 & 0.8188 & 0.7837 \\
& & MTL-HPS       & 0.9194 & 0.6015 & \textbf{0.8378} & 0.7862 \\
& & MTL-SA (Ours) & \textbf{0.9234} & \textbf{0.6202} & 0.8362 & \textbf{0.7933} \\
\cmidrule{2-7}
& \multirow{3}{*}{MoLFormer}
& STL           & 0.9429 & 0.6239 & 0.8367 & 0.8012 \\
& & MTL-HPS       & 0.9390 & 0.6321 & 0.8475 & 0.8062 \\
& & MTL-SA (Ours) & \underline{\textbf{0.9482}} & \textbf{0.6438} & \underline{\textbf{0.8528}} & \underline{\textbf{0.8149}} \\
\bottomrule
\end{tabular}
}
\end{table}

\subsubsection{Regularization Ablation}

Table~\ref{tab:ablation} evaluates L1 regularization strength $\lambda$ using the MoLFormer backbone with fixed hyperparameters to isolate the effect of sparsity regularization. Contrary to the concern that imposing sparsity might overly constrain feature selection and degrade accuracy, applying an appropriate regularization strength ($\lambda=10^{-3}$) in our framework yields dual benefits: it facilitates model interpretability, while simultaneously improving predictive performance relative to the unregularized baseline ($\lambda=0$). On ClinTox, SIDER and Tox21, moderate regularization achieves gains of +0.36\%, +0.23\% and +0.38\% respectively. 

\begin{table}[htbp]
\centering
\caption{
    \textbf{Ablation Study on L1 Sparsity Regularization Strength ($\lambda$).} ROC-AUC performance with MoLFormer is reported for varying values of the hyperparameter $\lambda$. The percentage change ($\Delta$) is calculated relative to the unregularized baseline ($\lambda$ = 0). }
\label{tab:ablation}
\footnotesize
\setlength{\tabcolsep}{3pt}
\begin{adjustbox}{max width=250pt}
\begin{tabular}{ccccccc}
\toprule
& \multicolumn{2}{c}{ClinTox} & \multicolumn{2}{c}{SIDER} & \multicolumn{2}{c}{Tox21} \\
$\lambda$ & ROC-AUC & $\Delta$ (\%) & ROC-AUC & $\Delta$ (\%) & ROC-AUC & $\Delta$ (\%) \\
\midrule
0    & 0.9439 & --      & 0.6184 & --      & 0.8395 & --      \\
\midrule
$10^{-4}$ & 0.9427 & -0.12\% & 0.6194 & +0.16\% & 0.8393 & -0.02\% \\
\midrule
$10^{-3}$  & 0.9473 & +0.36\% & 0.6198 & +0.23\% & 0.8427 & +0.38\% \\
\midrule
$10^{-2}$   & 0.9464 & +0.26\% & 0.6169 & -0.24\% & 0.8412 & +0.20\% \\
\bottomrule
\end{tabular}
\end{adjustbox}
\end{table}

\subsection{Interpretability via Attention Patterns}

The tokenization strategy of MoLFormer inherently supports atom-level analysis~\cite{ross2022large}. We leverage this to examine whether our sparsity constraint successfully isolates chemically relevant fragments. By visualizing the learned attention masks, we observe that the model identifies chemically interpretable features rather than relying on diffuse or irrelevant background context. We illustrate this capability using examples from the Tox21 and SIDER datasets, which effectively demonstrate two key behaviors: task-specific differentiation and cross-scaffold consistency. 

\begin{figure}
    \includegraphics[width = 0.91\linewidth]{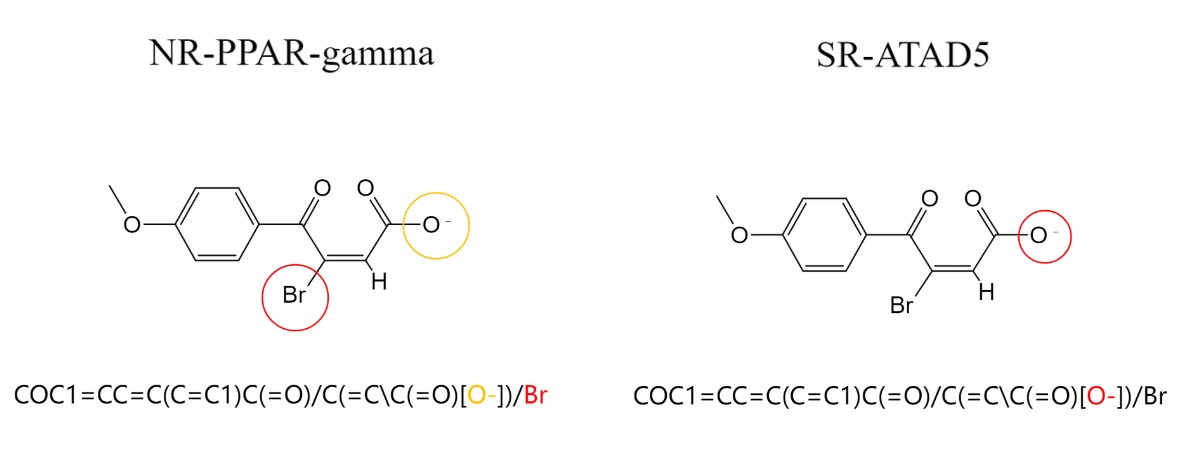}
    \centering
    \caption{
        \textbf{Task-Specific Differentiation on Tox21.}
        The model isolates distinct features within the same molecule for different endpoints. For the NR-PPAR-gamma task (left), attention targets the bromine (Br) atom, whereas for the SR-ATAD5 task (right), it focuses on the anionic oxygen (O$^-$). Salient fragments with high attention weights are highlighted in red. 
    }
    \label{fig:tox21_example}
\end{figure}

First, the model demonstrates the ability to attend to distinct features within the same molecule depending on the target toxicity endpoint. As shown in Fig.~\ref{fig:tox21_example}, for the NR-PPAR-gamma task (left), the attention mechanism heavily weights the bromine (Br) atom on the vinyl fragment. In contrast, for the SR-ATAD5 task (right), attention shifts exclusively to the anionic oxygen (O$^-$) of the carboxylate group. This differentiation suggests that the task-specific gating mechanism effectively filters the shared representation, allowing each head to focus on specific atomic features deemed most relevant to its biological endpoint. 

\begin{figure}
    \includegraphics[width = 0.91\linewidth]{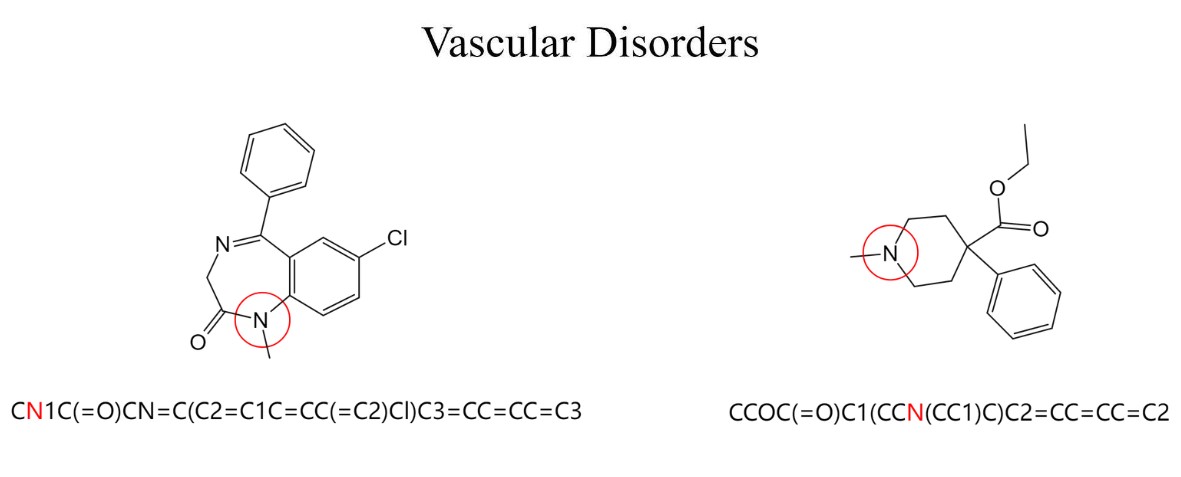}
    \centering
    \caption{
        \textbf{Toxicophore Generalization across Scaffolds on SIDER.}
        For the Vascular Disorders task, the model consistently identifies nitrogen-based motifs across structurally diverse drugs. It highlights the amide nitrogen in Diazepam (left) and the piperidine nitrogen in Meperidine (right), suggesting the learning of abstract chemical principles rather than rote memorization. 
    }
    \label{fig:sider_example}
\end{figure}

Second, the model exhibits consistency in feature selection across structurally diverse molecules, a property essential for robust generalization. In the SIDER Vascular Disorders task (Fig.~\ref{fig:sider_example}), the model consistently highlights nitrogen atoms despite significant structural differences. Specifically, it isolates the amide nitrogen within the benzodiazepine ring of Diazepam (left) and the piperidine nitrogen within the phenylpiperidine scaffold of Meperidine (right). This indicates that the sparsity penalty constrains the model to prioritize common nitrogen-based motifs shared among active compounds, rather than memorizing specific molecular scaffolds. 

These visualizations indicate that the L1 sparsity penalty does not merely reduce noise, but actively encourages the model to learn generalizable structural patterns. By isolating a minimal set of highly weighted tokens, the framework provides a transparent basis for its predictions, allowing domain experts to assess the chemical plausibility of the decision-making process. 

\section{Conclusions}

In this work, we introduced a MTL framework that enhances molecular toxicity prediction by integrating task-specific sparse attention. Validated across multiple MoleculeNet benchmarks, our approach consistently outperforms both single-task and standard MTL baselines. Crucially, our findings demonstrate that imposing moderate L1 sparsity regularization can simultaneously improve predictive accuracy and model interpretability, challenging the conventional view of a trade-off between these objectives. The learned attention patterns highlight salient molecular fragments, exhibiting task-specific discrimination and cross-molecule generalization. By isolating these discrete structural motifs, the framework offers an interpretable rationale for its predictions, enabling validation of whether the model's focus aligns with chemical intuition. Future work could explore explicit task-weighting strategies to further mitigate negative transfer and extend the framework to tackle large-scale, heterogeneous data problems~\cite{rahman2018processing}. By bridging high-performance prediction with transparent feature selection, our approach represents a promising step towards more reliable and interpretable \textit{in silico} toxicology. 

\bibliographystyle{IEEEtran}
\bibliography{refs}

@article{chithrananda2020chemberta,
  title={ChemBERTa: large-scale self-supervised pretraining for molecular property prediction},
  author={Chithrananda, Seyone and Grand, Gabriel and Ramsundar, Bharath},
  journal={arXiv preprint arXiv:2010.09885},
  year={2020}
}

@article{ross2022large,
  title={Large-scale chemical language representations capture molecular structure and properties},
  author={Ross, Jerret and Belgodere, Brian and Chenthamarakshan, Vijil and Padhi, Inkit and Mroueh, Youssef and Das, Payel},
  journal={Nature Machine Intelligence},
  volume={4},
  number={12},
  pages={1256--1264},
  year={2022},
  publisher={Nature Publishing Group UK London}
}

@inproceedings{wang2019smiles,
  title={Smiles-bert: large scale unsupervised pre-training for molecular property prediction},
  author={Wang, Sheng and Guo, Yuzhi and Wang, Yuhong and Sun, Hongmao and Huang, Junzhou},
  booktitle={Proceedings of the 10th ACM international conference on bioinformatics, computational biology and health informatics},
  pages={429--436},
  year={2019}
}

@article{caruana1997multitask,
  title={Multitask learning},
  author={Caruana, Rich},
  journal={Machine learning},
  volume={28},
  number={1},
  pages={41--75},
  year={1997},
  publisher={Springer}
}

@article{ramsundar2015massively,
  title={Massively multitask networks for drug discovery},
  author={Ramsundar, Bharath and Kearnes, Steven and Riley, Patrick and Webster, Dale and Konerding, David and Pande, Vijay},
  journal={arXiv preprint arXiv:1502.02072},
  year={2015}
}

@article{allenspach2024neural,
  title={Neural multi-task learning in drug design},
  author={Allenspach, Stephan and Hiss, Jan A and Schneider, Gisbert},
  journal={Nature Machine Intelligence},
  volume={6},
  number={2},
  pages={124--137},
  year={2024},
  publisher={Nature Publishing Group UK London}
}

@article{peng2020empirical,
  title={An empirical study of multi-task learning on BERT for biomedical text mining},
  author={Peng, Yifan and Chen, Qingyu and Lu, Zhiyong},
  journal={arXiv preprint arXiv:2005.02799},
  year={2020}
}

@article{zhang2022pushing,
  title={Pushing the boundaries of molecular property prediction for drug discovery with multitask learning BERT enhanced by SMILES enumeration},
  author={Zhang, Xiao-Chen and Wu, Cheng-Kun and Yi, Jia-Cai and Zeng, Xiang-Xiang and Yang, Can-Qun and Lu, Ai-Ping and Hou, Ting-Jun and Cao, Dong-Sheng},
  journal={Research},
  volume={2022},
  pages={0004},
  year={2022},
  publisher={AAAS}
}

@inproceedings{caruna1993multitask,
  title={Multitask learning: A knowledge-based source of inductive bias},
  author={Caruna, Rich},
  booktitle={Machine learning: Proceedings of the tenth international conference},
  pages={41--48},
  year={1993}
}

@article{zhu2021genotype,
  title={Genotype-determined EGFR-RTK heterodimerization and its effects on drug resistance in lung Cancer treatment revealed by molecular dynamics simulations},
  author={Zhu, Mengxu and Wang, Debby D and Yan, Hong},
  journal={BMC molecular and cell biology},
  volume={22},
  number={1},
  pages={34},
  year={2021},
  publisher={Springer}
}

@article{ruder2017overview,
  title={An overview of multi-task learning in deep neural networks},
  author={Ruder, Sebastian},
  journal={arXiv preprint arXiv:1706.05098},
  year={2017}
}

@inproceedings{liu2019end,
  title={End-to-end multi-task learning with attention},
  author={Liu, Shikun and Johns, Edward and Davison, Andrew J},
  booktitle={Proceedings of the IEEE/CVF conference on computer vision and pattern recognition},
  pages={1871--1880},
  year={2019}
}

@article{wu2018moleculenet,
  title={MoleculeNet: a benchmark for molecular machine learning},
  author={Wu, Zhenqin and Ramsundar, Bharath and Feinberg, Evan N and Gomes, Joseph and Geniesse, Caleb and Pappu, Aneesh S and Leswing, Karl and Pande, Vijay},
  journal={Chemical science},
  volume={9},
  number={2},
  pages={513--530},
  year={2018},
  publisher={Royal Society of Chemistry}
}

@article{gayvert2016data,
  title={A data-driven approach to predicting successes and failures of clinical trials},
  author={Gayvert, Kaitlyn M and Madhukar, Neel S and Elemento, Olivier},
  journal={Cell chemical biology},
  volume={23},
  number={10},
  pages={1294--1301},
  year={2016},
  publisher={Elsevier}
}

@article{kuhn2016sider,
  title={The SIDER database of drugs and side effects},
  author={Kuhn, Michael and Letunic, Ivica and Jensen, Lars Juhl and Bork, Peer},
  journal={Nucleic acids research},
  volume={44},
  number={D1},
  pages={D1075--D1079},
  year={2016},
  publisher={Oxford University Press}
}

@article{huang2016modelling,
  title={Modelling the Tox21 10 K chemical profiles for in vivo toxicity prediction and mechanism characterization},
  author={Huang, Ruili and Xia, Menghang and Sakamuru, Srilatha and Zhao, Jinghua and Shahane, Sampada A and Attene-Ramos, Matias and Zhao, Tongan and Austin, Christopher P and Simeonov, Anton},
  journal={Nature communications},
  volume={7},
  number={1},
  pages={10425},
  year={2016},
  publisher={Nature Publishing Group UK London}
}

@article{wolf2019huggingface,
  title={Huggingface's transformers: State-of-the-art natural language processing},
  author={Wolf, Thomas and Debut, Lysandre and Sanh, Victor and Chaumond, Julien and Delangue, Clement and Moi, Anthony and Cistac, Pierric and Rault, Tim and Louf, R{\'e}mi and Funtowicz, Morgan and others},
  journal={arXiv preprint arXiv:1910.03771},
  year={2019}
}

@article{sanh2019distilbert,
  title={DistilBERT, a distilled version of BERT: smaller, faster, cheaper and lighter},
  author={Sanh, Victor and Debut, Lysandre and Chaumond, Julien and Wolf, Thomas},
  journal={arXiv preprint arXiv:1910.01108},
  year={2019}
}

@inproceedings{sechidis2011stratification,
  title={On the stratification of multi-label data},
  author={Sechidis, Konstantinos and Tsoumakas, Grigorios and Vlahavas, Ioannis},
  booktitle={Joint European conference on machine learning and knowledge discovery in databases},
  pages={145--158},
  year={2011},
  organization={Springer}
}

@inproceedings{akiba2019optuna,
  title={Optuna: A next-generation hyperparameter optimization framework},
  author={Akiba, Takuya and Sano, Shotaro and Yanase, Toshihiko and Ohta, Takeru and Koyama, Masanori},
  booktitle={Proceedings of the 25th ACM SIGKDD international conference on knowledge discovery \& data mining},
  pages={2623--2631},
  year={2019}
}

@article{loshchilov2017decoupled,
  title={Decoupled weight decay regularization},
  author={Loshchilov, Ilya and Hutter, Frank},
  journal={arXiv preprint arXiv:1711.05101},
  year={2017}
}

@article{rahman2018processing,
  title={Processing of electronic medical records for health services research in an academic medical center: methods and validation},
  author={Rahman, Nabilah and Wang, Debby D and Ng, Sheryl Hui-Xian and Ramachandran, Sravan and Sridharan, Srinath and Khoo, Astrid and Tan, Chuen Seng and Goh, Wei-Ping and Tan, Xin Quan},
  journal={JMIR Medical Informatics},
  volume={6},
  number={4},
  pages={e10933},
  year={2018},
  publisher={JMIR Publications Inc., Toronto, Canada}
}

\end{document}